\pdfoutput=1

\documentclass[%
 reprint,
 superscriptaddress,
 nobibnotes,
 amsmath,amssymb,
 aps,
 showkeys,
 longbibliography,
]{revtex4-2}

\usepackage{graphicx}
\usepackage{dcolumn}
\usepackage{bm}
\usepackage{chemformula}
\usepackage[separate-uncertainty=true,separate-uncertainty,multi-part-units=single]{siunitx}
\usepackage{hyperref}
\hypersetup{colorlinks=true,allcolors=magenta}
\usepackage{booktabs}

\begin{document}

\title{Symmetry-Informed Graph Neural Networks for Carbon Dioxide Isotherm and Adsorption Prediction in Aluminum-Substituted Zeolites}

\author{Marko Petkovi\'{c}}
    \affiliation{Materials Simulation and Modelling, Department of Applied Physics and Science Education, Eindhoven University of Technology, Eindhoven}
    \affiliation{Eindhoven Artificial Intelligence Systems Institute, Eindhoven University of Technology, Eindhoven }
\author{Jos{\'e}-Manuel Vicent Luna}
    \email[Corresponding author: ]{j.vicent.luna@tue.nl}
    \affiliation{Materials Simulation and Modelling, Department of Applied Physics and Science Education, Eindhoven University of Technology, Eindhoven}
\author{El\={\i}za Beate Dinne}
    \affiliation{Materials Simulation and Modelling, Department of Applied Physics and Science Education, Eindhoven University of Technology, Eindhoven}
\author{Vlado Menkovski }
    \affiliation{Eindhoven Artificial Intelligence Systems Institute, Eindhoven University of Technology, Eindhoven }
    \affiliation{Data and AI, Department of Mathematics and Computer Science, Eindhoven University of Technology, Eindhoven}
\author{Sof\'{i}a Calero}
    \email[Corresponding author: ]{s.calero@tue.nl}
    \affiliation{Materials Simulation and Modelling, Department of Applied Physics and Science Education, Eindhoven University of Technology, Eindhoven}
    \affiliation{Eindhoven Artificial Intelligence Systems Institute, Eindhoven University of Technology, Eindhoven }


\begin{abstract}

Accurately predicting adsorption properties in nanoporous materials using Deep Learning models remains a challenging task. This challenge becomes even more pronounced when attempting to generalize to structures that were not part of the training data.. In this work, we introduce SymGNN, a graph neural network architecture that leverages material symmetries to improve adsorption property prediction. By incorporating symmetry operations into the message-passing mechanism, our model enhances parameter sharing across different zeolite topologies, leading to improved generalization. We evaluate SymGNN on both interpolation and generalization tasks, demonstrating that it successfully captures key adsorption trends, including the influence of both the framework and aluminium distribution on CO$_2$ adsorption. Furthermore, we apply our model to the characterization of experimental adsorption isotherms, using a genetic algorithm to infer likely aluminium distributions. Our results highlight the effectiveness of machine learning models trained on simulations for studying real materials and suggest promising directions for fine-tuning with experimental data and generative approaches for the inverse design of multifunctional nanomaterials.

\end{abstract}


\maketitle


\section{Introduction}
\label{sec:intro}
In recent years, there has been a noticeable increase in atmospheric CO$_2$ levels, with the corresponding rise in greenhouse effects, highlighting the pressing need for effective carbon mitigation strategies. Carbon capture emerges as a viable approach to address this issue \cite{odunlami2022advanced}, and nanoporous materials, specifically zeolites, stand out as promising candidates \cite{kumar2020utilization}. Zeolites exhibit a notable capacity for gas adsorption, making them well-suited for reducing carbon levels in the atmosphere. This capacity is commonly analyzed through adsorption isotherms, which describe how the amount of CO$_2$ adsorbed varies with pressure and provide insights into the material's efficiency and suitability for carbon capture. Their appeal extends further with attributes such as high thermal stability \cite{cruciani2006zeolites} and cost-effectiveness in synthesis when compared to other adsorbents \cite{khaleque2020zeolite}. 

Additionally, the extensive variety of synthesizable zeolite topologies \cite{derbe2021short}, each characterized by distinct pore sizes and properties, adds a layer of versatility to their application. Within a zeolite topology, there are multiple possible configurations, as a result of different silicon and aluminium atom arrangements. These configurations can have different CO$_2$ adsorption properties, where the overall trend is that an increase in aluminium atoms leads to better adsorption properties \cite{romero2023adsorption}. However, for the same Si/Al ratio there can still be a considerable variance in properties such as the heat of adsorption and the adsorption isotherms.

Due to the large configuration space of possible zeolite topologies and Si/Al configurations, experimentally studying each configuration to find structures with desirable properties is impossible. In this context, simulations provide a powerful alternative, enabling the prediction of adsorption properties without the need for extensive synthesis and testing \cite{kim2012efficient,pham2014experimental,kellouai2022gas,jeong2016understanding,fischer2013modeling,krishna2010silico}. However, certain computational methods, particularly classical simulations such as Grand Canonical Monte Carlo (GCMC), require sampling at multiple pressures to generate adsorption isotherms and fully characterize a material's adsorption behavior. This can be computationally expensive, making it challenging to efficiently screen large numbers of candidate structures.

To this end, Deep Learning (DL) can be a powerful tool for accelerating the discovery and characterization of materials \cite{jablonka2020big,liu2025comprehensive,choudhary2022recent,reiser2022graph}. For predicting the properties of crystals, several Graph Neural Network (GNN) architectures \cite{xie2018crystal,schutt2018schnet,chen2019graph,choudhary2021atomistic,ruff2024connectivity} and Transformer-based models \cite{yan2022periodic,taniai2024crystalformer} have been proposed, which operate on atomic types and positions within the unit cell. In addition, generative models have been increasingly explored for the design of novel materials, allowing the discovery of structures with targeted properties \cite{xie2021crystal,jiao2024crystal,miller2024flowmm,levy2024symmcd}. Furthermore, various DL approaches have been specifically tailored for nanoporous materials, such as zeolites and Metal-Organic Frameworks (MOFs). Some of these methods focus on predicting adsorption behavior across different adsorbates \cite{wang2020accelerating,liu2023zeonet,petkovic2024graph,kang2023multi,chen2022interpretable,cao2023moformer}, while others aim to design new materials with optimized adsorption and structural properties \cite{fu2023mofdiff,kang2024chatmof}.

Most of these models explicitly respect and leverage the symmetries present in a crystal by being invariant or equivariant to the Euclidean group \( E(3) \), as well as the periodic boundary conditions. Each crystal has an associated Space Group (SG), which is a subgroup of \( E(3) \) and determines the equivalent atomic positions within the unit cell. By incorporating this information, geometric constraints can be directly embedded into the neural network architecture. Although several approaches for predicting crystal properties account for space group information, they either neglect symmetries at the unit cell level \cite{kaba2022equivariant} or lack generalizability across materials with different topologies \cite{petkovic2023equivariant}. These approaches introduce separate parameters in the GNN for the node and message update functions for nodes/edges, which are considered symmetrically equivalent. 

In this work, we introduce SymGNN, a symmetry-informed graph neural network architecture designed to incorporate crystal symmetries into message passing. By leveraging symmetry operations, our model enables more effective parameter sharing across different zeolite topologies, leading to improved generalization. We demonstrate that SymGNN successfully predicts both adsorption isotherms and heats of adsorption for unseen topologies, capturing key adsorption trends by effectively modeling the influence of both the framework structure and the Si/Al distribution on adsorption properties. Finally, we show that our model can be applied to characterize experimental adsorption isotherms by inferring structural properties such as the Si/Al ratio, potentially enhancing materials charaterization and analysis.
\section{Crystal Symmetries}
\subsection{Unit Cell}
In crystalline materials, the arrangement of atoms follows a repeating periodic structure, which is described using the Bravais lattice $\Lambda$. A Bravais lattice defines the periodic arrangement of points in space, and the structure of the entire crystal can be generated by translating these points along the lattice vectors. Equation \ref{eq:bravais} describes the Bravais lattice, where $\mathbf{a}_i$ are the linearly independent basis vectors of the lattice and $m_i$ are their integer multiples. This defines the periodicity of the lattice in a three-dimensional space.

\begin{equation}
    \Lambda = \left\{ \sum_i^3 m_i \mathbf{a}_i\ |\ m_i \in \mathbb{Z}\right\} \label{eq:bravais}
\end{equation}

From the Bravais lattice, we can define the unit cell $U$, which represents the smallest repeating unit in the crystal structure. The unit cell can be defined using the basis vectors of the crystal lattice, as shown in Equation \ref{eq:unitcell}. Here, $x_i$ are the fractional coordinates of the points in space belonging to the unit cell.

\begin{equation}
    U = \left\{ \sum_i^3 x_i \mathbf{a}_i\ |\  0 \leq x_i < 1 \right\} \label{eq:unitcell}
\end{equation} 

The set of atoms $S$ contained within a unit cell is defined by Equation \ref{eq:unitcellatoms}, in which $Z_i$ is the atomic number, and $\mathbf{x}_i$ is the position in fractional coordinates of an atom. By combining the bravais lattice and the set of atoms in the unit cell, we can fully describe the crystal structure.

\begin{equation}
    S = \left\{ (Z_i, \mathbf{x}_i)\ |\  \mathbf{x}_i \in U \right\} 
    \label{eq:unitcellatoms}
\end{equation}

\subsection{Space Group}
Crystals exhibit a high degree of symmetry, which plays a crucial role in determining their physical properties. The symmetry of a crystal can be described mathematically by a space group $G$. A space group encompasses the full set of symmetry operations that can be applied to the crystal, leaving it invariant. As such, it captures all of the rotational, reflectional, and translational symmetries of the structure. 

Each element of the space group is a group action $g$. Each group action consists of a tuple of a linear transformation $\mathbf{W}$ and a translation vector $\mathbf{t}$. The elements of a space group act on a position $\mathbf{x}$ as shown in Equation \ref{eq:groupaction}. 
\begin{equation}
g \cdot \mathbf{x} =  \mathbf{Wx} + \mathbf{t} \label{eq:groupaction}
\end{equation}

One important property of space groups is their closure under multiplication. This means that when two elements of the space group are multiplied, the result is another element of the same space group. This closure property is described by Equations \ref{eq:sg_mulW} and \ref{eq:sg_mult}.
\begin{align}
    \mathbf{W}' &= \mathbf{W}_1 \mathbf{W}_2 \label{eq:sg_mulW} \\
    \mathbf{t}' &= \mathbf{W}_2 \mathbf{t}_1 + \mathbf{t_2} \label{eq:sg_mult}
\end{align}

\subsection{Group Orbit}
The orbit of an atom is the set of all positions which the atom can be mapped to by elements of the space group, and can be formally defined using Equation \ref{eq:orbit}. Atoms that belong to the same orbit are considered to be equivalent under the space group symmetry. The cardinality (or size) of an atom's orbit depends on its position within the crystal. Specifically, an atom located in the least symmetric position will have an orbit that includes all the space group operations, meaning its orbit will have the same cardinality as the space group. In contrast, an atom in a more symmetric position will have a smaller orbit, as some space group operations may map the atom to equivalent positions within the unit cell, reducing the total number of distinct positions in the orbit.
\begin{equation}
    \text{Orbit}(\mathbf{x}) = \left\{g \cdot \mathbf{x} \ |\ g \in G\right \} \label{eq:orbit}
\end{equation}

Next, we will define the set of operations that can map each position in an orbit to every other position, except the original position. For orbits with the same cardinality as the space group, this set will coincide with the full set of space group operations, minus the identity operation. However, for smaller orbits (those with fewer positions), some of the space group operations may be redundant as they do not contribute to mapping positions within the orbit. In such cases, the set of operations that maps one position to another will be a proper subset of the full space group. Mathematically, this set of operations is defined in Equation \ref{eq:orbitops}.
\begin{equation}
    \text{Ops}(\mathbf{x}) = \left \{ g \in G | g \cdot \mathbf{x} \in \text{Orbit}(\mathbf{x}) \wedge g \cdot \mathbf{x} \neq \mathbf{x} \right \} \label{eq:orbitops}
\end{equation}

\subsection{Generators}
To define the generators of the set of operations associated with an orbit, we need to identify the minimal set of operations that, when combined (with repetition) through multiplication, can generate all other operations that map positions within the orbit. These generators are crucial because they form the core operations that preserve the symmetry of the crystal while minimizing redundancy.

Mathematically, we define the set of generators, $\text{Gen}(\text{Ops}(\mathbf{x}))$, as the minimal subset of operations (Equation \ref{eq:gensub}) such that every operation in $\text{Ops}(\mathbf{x})$ can be expressed as a product of elements from this set (Equation \ref{eq:gengen}). This set of generators can be thought of as the building blocks for the full set of orbit operations.
\begin{align}
    \text{Gen}(\text{Ops}(\mathbf{x})) &\subseteq \text{Ops}(\mathbf{x}) \label{eq:gensub}\\
    \left< \text{Gen}(\text{Ops}(\mathbf{x}))  \right> &= \text{Ops}(\mathbf{x}) \label{eq:gengen}
\end{align}

In this equation, $\left < S \right > $ denotes the subgroup generated by the set $S$. As such, every element $g \in \text{Ops}(\mathbf{x})$ can be defined using the generators, as shown in Equation \ref{eq:genops}. 
\begin{equation}
    g = g_i^{n_1 } g_2^{n_2}...g_k^{n_k}, \quad n_i \in \mathrm{Z}, \quad g_1, g_2, ...,g_k  \in \text{Gen}(\text{Ops}(\mathbf{x})) \label{eq:genops}
\end{equation}

However, there can still be multiple minimal yet distinct sets of generators for a given set of symmetry operations. For example, in the cyclic group of order 4 ($C_4$), both a 90-degree rotation and a 270-degree rotation can independently generate all other elements of $C_4$. To ensure a consistent choice of generators for a given position $\mathbf{x}$, we adopt the generator sets defined for different space groups as provided by the Bilbao Crystallographic Server (BCS) \cite{aroyo2006bilbao}.

\section{Methods}
\subsection{Zeolite Frameworks}
For this work, we used 106 different zeolite topologies with varying structural features. For each topology, varying configurations of silicon and aluminium atoms were generated, with the lowest Si/Al ratio being 3. The different configurations for each topology were generated using the ZEORAN\cite{romero2023adsorption} program and the PORRAN program, which is a Python extension of ZEORAN. These programs make use of four different algorithms to place aluminium atoms in an all-silica zeolite. These algorithms place the aluminium atoms either in \textit{clusters}, \textit{chains}, uniformly (\textit{maximum entropy}) or \textit{randomly}.  Depending on the algorithm, the Lowenstein rule might be broken (Al-O-Al), since recent studies \cite{afeworki2004synthesis,pavon2014direct,fletcher2017violations,heard2019effect} found structures breaking this rule. A more detailed description of the algorithms can be found in the SI\dag. Si/Al configurations for the MOR, RHO, MFI and ITW were taken from \citet{petkovic2024graph}. For the other structures, atomic coordinates for pure silica were taken from IZA \cite{baerlocher2007atlas}, following which Si/Al configurations were generated using the aforementioned algorithms. In total, 27648 structures were generated.

\subsection{Computational Details}
In this study, we investigated the CO$_2$ adsorption isotherm and heat of adsorption ($-\Delta H$). These properties can give us insight into the CO$_2$ adsorption in zeolites. The heat of adsorption can give an indication about the interaction strength between the zeolite and the adsorbate, whereas the isotherm can tell us about the adsorption capacity of a zeolite at different pressures.
To calculate the heat of adsorption, simulations using the Widom particle insertion method in the canonical ensemble ($NVT$) were performed \cite{widom1963some} for 200,000 cycles. For the CO$_2$ adsorption isotherms, simulations were carried out using the grand canonical ensemble ($\mu VT$), where the loading was calculated for a range of pressures between 0.01 kPA and 10,000 kPA.

 The isotherms were calculated for the MOR, MFI, MEL, TON, and ITW zeolites. To obtain an adsorption isotherm for a single Si/Al configuration of a zeolite, multiple simulations need to be carried out. To generate a large dataset of adsorption isotherms efficiently, some simulations were sped up by using a reduced number of unit cells, depending on the zeolite. We validated this approximation by comparing isotherms varying the Si/Al ratio using full (i.e., the number of unit cells ensures that the simulation box is longer than twice the cutoff in each direction) and reduced simulation boxes of each zeolite. We found that the number of unit cells can be reduced for MOR, MFI, and MEL, without compromising the adsorption results. However, using the reduced simulation box, we found more fluctuations for TON and ITW. Therefore, we use the full simulation box for these two zeolites. The full and reduced number of unit cells and the verification procedure and results of the verification can be found in the SI\dag. Finally, we fitted the 2-site Langmuir-Freundlich model (Equation \ref{eq:2slf}) using RUPTURA \cite{sharma2023ruptura}, which can smooth out possible fluctuations as a consequence of using reduced simulation boxes. 

\begin{equation}
    q(p) = \sum_i^2 q_i^{\text{sat}} \frac{b_ip^{\nu_i}}{1+b_ip^{\nu_i}} \label{eq:2slf}
\end{equation}

The RASPA software \cite{dubbeldam2016raspa} was used to carry out all the simulations. The force field and point charges used for the simulations were taken from \citet{romero2023adsorption}. It extends the force field introduced in \citet{garcia2009transferable}, by accounting for atoms breaking the L{\"o}wenstein rule. For each zeolite configuration, sodium cations were introduced to balance the difference in charge as a result of the aluminium substitutions. The simulations were carried out at room temperature (298K). 

\subsection{Dataset}\label{sec:dataset}

\begin{figure}[h]
\centering
  \includegraphics[width=.9\linewidth]{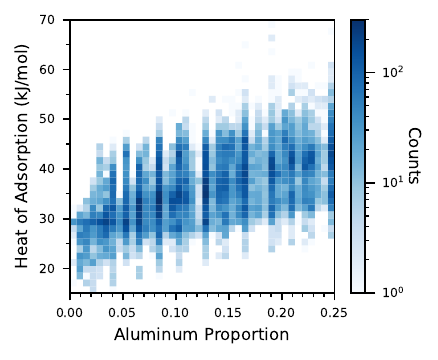}
  \caption{Heat of adsorption for all datapoints as a function of the aluminium proportion. Note that the color is in log-scale.}
  \label{fig:hoadist}
\end{figure}

In Figure \ref{fig:hoadist}, the relationship between the proportion of aluminium atoms and the heat of adsorption is visualized. Overall, there is a slight trend for an increasing heat of adsorption with a higher aluminium proportion. However, there is still a significant dependence of the heat of adsorption on both the framework type, as well as the distribution of aluminium atoms within the framework. Sodium cations have been shown to reside close to the aluminium framework atoms \cite{romero2023adsorption}, and can thus affect the strength of adsorption sites. Furthermore, the geometry of the framework pores also plays a role in the adsorption strength.

\begin{figure}[h]
\centering
  \includegraphics[width=.9\linewidth]{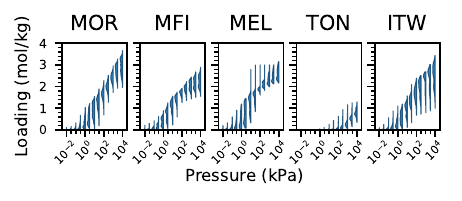}
  \caption{Distribution of loading values at each simulated pressure.}
  \label{fig:isoviolin}
\end{figure}

\begin{figure*}[]
\centering
  \includegraphics[width=.95\linewidth]{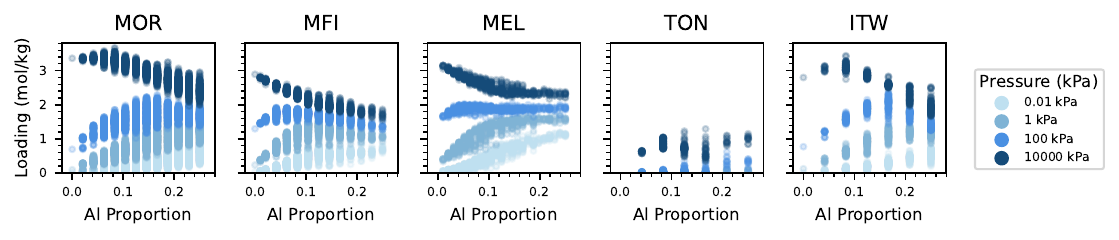}
  \caption{Loading values for all datapoints with isotherms as a function of the aluminium proportion, at varying pressures.}
  \label{fig:isoscatter}
\end{figure*}

Similarly, the behaviour of the adsorption isotherms is also impacted by the aluminium distribution and the geometry of the material. As can be seen in Figure \ref{fig:isoviolin}, the shape of the isotherms can vary greatly between topologies, showing how the geometry of the pores plays a role in the isotherm. Furthermore, there is a significant variance in the isotherms for the same zeolite topology, suggesting that the distribution and ratio of aluminium atoms plays a role. This can be seen in Figure \ref{fig:isoscatter}, where the loading for a given pressure and aluminium proportion is shown for each zeolite topology. In general, when increasing the pressure, the loading first increases for structures with a higher aluminium proportion. However, at higher pressures, these structures tend to reach saturation earlier, whereas structures with a lower aluminium proportion tend to achieve a higher loading.

Using this data, we define two different splits of the data. In the first split, the \textit{generalization} split, the model is evaluated on the ITW and CHA structures, and trained on the remaining zeolites. As such, the model will not have seen the structure of ITW and CHA. Therefore, we can use this test set to evaluate how well the model has learned how the structure and distribution of aluminium atoms of a zeolite impact its adsorption properties. In the second split, \textit{interpolation} split, the data is split in training, validation and testing set. For each zeolite, the different configurations are split in an 80:10:10 between the three sets. Using this test set, we can evaluate how well the model understands the effect of the aluminium distribution within each topology. 

In our dataset, there is a large class imbalance, with MOR having 4300 structures present in the dataset, and EUO having only 78. To avoid the model overfitting on more prevalent structures, we over- and under-sample the configurations of different zeolites, to ensure the model has seen 250 structures per zeolite during an epoch. The number of structures for each zeolite topology can be found in the SI\dag.

\section{Symmetry-Informed Graph Neural Networks}
Several existing GNN architectures \cite{kaba2022equivariant,petkovic2023equivariant} have leveraged crystal symmetries to enhance their performance. These models make use of symmetry-based parameter sharing, where unique node and message update functions are assigned to each set of equivalent nodes and edges that belong to the same orbit. This approach increases the model's expressiveness, as a distinct set of parameters is learned for each (abstract) spatial relationship. This is analogous to how a convolutional neural network learns separate parameters for each pixel within a kernel.

However, when trained on a specific set of topologies, these models generally cannot be transferred to a new topology due to the lack of a clear mapping between sets of atomic orbits in different crystals. In \citet{kaba2022equivariant}, this challenge was addressed by defining symmetries between unit cells, allowing the model to be fully transferable. This was achieved by constructing a 2×2×2 supercell, which enabled the model to recognize equivalent relationships across unit cells. However, symmetries within the unit cell itself were not explicitly leveraged, meaning the approach does not take full advantage of all available symmetry information. As a result, while the model generalizes across different crystal topologies, it may not be as efficient or expressive as a model that fully incorporates intra-unit-cell symmetries.

\subsection{Symmetry-Informed Message Passing}
\begin{figure*}[]
\centering
  \includegraphics[width=.95\linewidth]{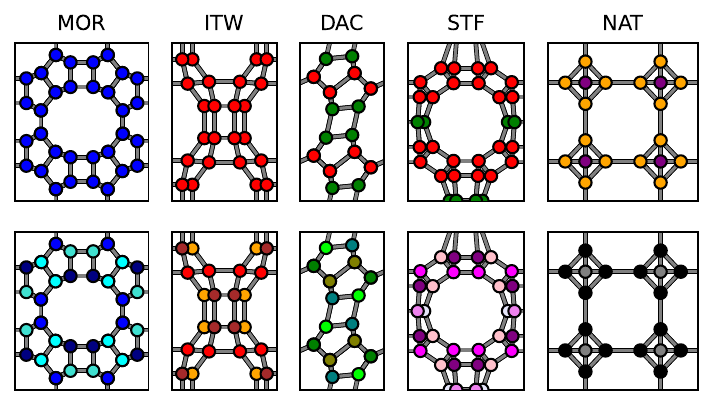}
  \caption{Comparison of parameter sharing in symmetry-informed message passing (top row) and symmetry-based parameter sharing (bottom row) across five different zeolite topologies. In the top row, nodes with the same generators are assigned the same color, while in the bottom row, nodes with the same node-update parameters (belonging to the same orbit) share a color. Notably, while symmetry-based parameter sharing results in more distinct colors, symmetry-informed message passing allows certain generator sets to be shared across different zeolites, enabling better transferability.} 
  \label{fig:parshare}
\end{figure*}
To address these limitations, we introduce \textbf{Symmetry-Informed Message Passing}, which explicitly incorporates the generators of the set of symmetry operations into the node update function. By doing so, the model is directly informed about how symmetries act within a given structure, allowing it to distinguish between equivalent and non-equivalent atomic environments in a way that generalizes across different topologies. Unlike previous approaches, which either lack transferability or fail to fully utilize symmetry information, our approach ensures that the model can recognize and leverage shared symmetries while maintaining the flexibility to adapt to new crystal structures.

The overall message-passing scheme is defined in Equations \ref{eq:message}-\ref{eq:node}. Here, $\mathbf{h}_i^l$ represents the embedding of node $i$ at layer $l$, while $\mathbf{e}_{ij}$ denotes the embedding of the edge connecting nodes $i$ and $j$. The set of generators associated with node $i$, denoted as $G_i$, encodes the local symmetry properties of the structure. Each message $\mathbf{m}_{ij}^l$ is computed from neighboring nodes and edges using the message function $\phi_e$, while node embeddings are updated through $\phi_h$, the node update function. Unlike standard message-passing approaches, $\phi_h$ is explicitly conditioned on $G_i$, allowing it to capture symmetry-aware representations and adapt its updates based on the geometric context of each node.
\begin{align}
    \mathbf{m}_{ij}^l = \phi_e(\mathbf{h}_i^l, \mathbf{h}_j^l, \mathbf{e}_{ij}) \label{eq:message} \\
    \mathbf{m}_i^l = \frac{1}{|\mathcal{N}_i|} \sum_{j \in \mathcal{N}_i} \mathbf{m}_{ij}^l \label{eq:meesage_agg} \\
    \mathbf{h}_i^{l+1} = \phi_h(\mathbf{h}_i^l, \mathbf{m}_i^l | G_i) \label{eq:node}
\end{align}
To condition the node update layer on the generators, we utilize feature-wise linear modulation (FiLM) \cite{brockschmidt2020gnn}, as described in Equations \ref{eq:deepset} and \ref{eq:film}. In the first step, we apply a standard weight multiplication for the node update. Then, we introduce $\gamma$ and $\beta$, which allow the model to adjust the feature values based on the symmetry information of the node. These parameters act as dynamic scaling factors, enabling the model to emphasize or suppress features according to the symmetries inherent in the crystal structure. To compute $\gamma$ and $\beta$, we embed the set of generators using a DeepSets-inspired model \cite{zaheer2017deep}. Each element of the set of generators is represented by flattening its rotation matrix and concatenating it with the corresponding translation vector. This approach captures the relationships between the generators in a permutation-invariant manner and provides the necessary modulating parameters for the node update.
\begin{align}
    \gamma_i, \beta_i = \text{DeepSets}(G_i) \label{eq:deepset} \\
    \phi_h(\mathbf{h}_i^l, \mathbf{m}_i^l | G_i) = \gamma_i \odot W(\mathbf{h}_i^l \| \mathbf{m}_i^l) + \beta_i \label{eq:film}
\end{align}

Figure \ref{fig:parshare} compares the utilization of symmetries in symmetry-informed message passing and symmetry-based parameter sharing. While symmetry-based parameter sharing introduces a greater number of distinct parameters, these assignments are specific to each topology and cannot be transferred between zeolites. Consequently, a new model must be trained for each topology. In contrast, symmetry-informed message passing enables certain generator sets to be shared across different zeolites. Furthermore, even when generator sets differ, they may still contain common symmetry operations, further enhancing parameter transferability.

\begin{figure*}[]
\centering
  \includegraphics[width=.95\linewidth]{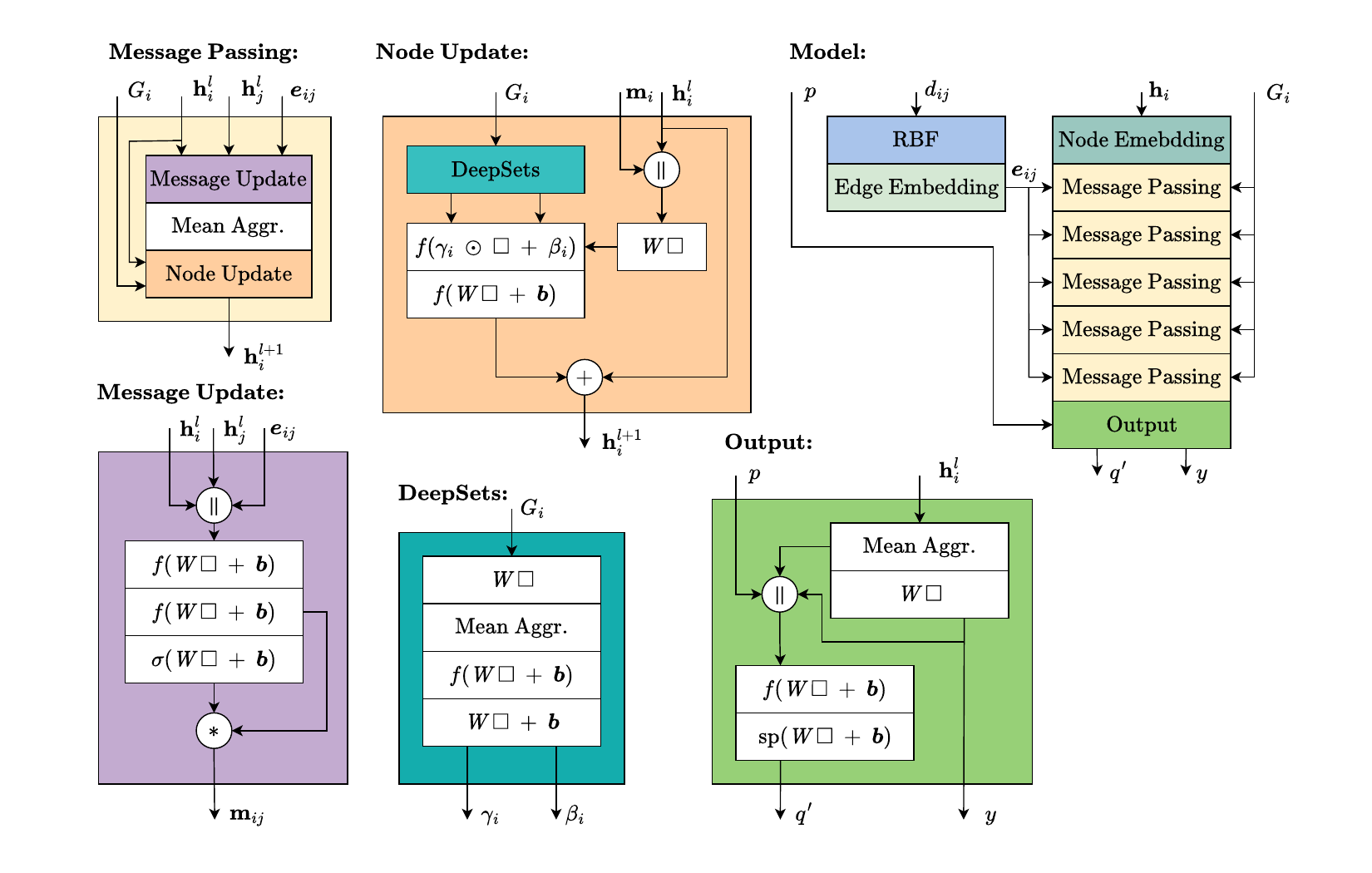}
  \caption{The SymGNN architecture. $\square$ denotes the layer input, $\|$ denotes concatenation and $\odot$ denotes elementwise multiplication. $f$ is the ELU activation function, $\sigma$ is the sigmoid activation and $\textbf{sp}$ is the Softplus activation. In the model, atoms and distances between atoms are embedded, following which symmetry-informed message passing takes places. In the output module, the final hidden state is used to predict the heat of adsorption ($y$). By combining the final hidden state, the predicted heat of adsorption and the pressure, the model predicts the derivative of the loading.}
  \label{fig:symgnn}
\end{figure*}
\subsection{Model Architecture}

To address the challenges of predicting adsorption properties in zeolites, we introduce \textbf{SymGNN}, a graph neural network that makes use of symmetry-informed message passing. This approach allows the model to efficiently predict the CO$_2$ heat of adsorption and adsorption isotherms across different zeolite structures by leveraging the inherent symmetries within the zeolite topologies. 

Since the adsorption isotherm is a function rather than a scalar, and is monotonically increasing with pressure, our model does not predict the loading at a given pressure directly. Instead, it predicts the derivative of the loading with respect to the pressure. Furthermore, rather than predicting the derivative of the loading at discrete pressures, our model predicts the isotherm function itself, similar to the approach used in neural operators. The model takes the final hidden state of the GNN, concatenates it with the pressure and predicted heat of adsorption, and passes it through a multi-layer perceptron (MLP) to produce the loading derivative predictions. To obtain the full isotherm for a given structure, the MLP is evaluated at different pressures. The resulting loading derivatives are then integrated to obtain the true loading. For numerical stability, both the calculation of the loading derivative and the integration process are performed with respect to the logarithm of the pressure. The precision of the predicted isotherm can be controlled by adjusting the number of pressures at which the MLP is evaluated. 

A full overview of the \textbf{SymGNN} architecture is provided in Figure \ref{fig:symgnn}. The model consists of 5 symmetry-informed message passing layers, each with hidden states of size 64. Nodes are embedded using a single linear layer, while edges are embedded using radial basis functions (RBF) \cite{schutt2018schnet} with 64 bins, followed by a linear layer. Messages are self-importance weighted, and aggregated using mean pooling. All linear layers in the message and node update steps are followed by layer normalization \cite{ba2016layer}. The DeepSets modules, which provide the parameters for FiLM in the node update, have an internal hidden state of 32. Throughout the model, the ELU activation function is used. To predict both the heat of adsorption and the loading derivative, mean aggregation     is used to obtain a graph-level representation, as adsorption properties are independent of the number of atoms in the unit cell.

\subsection{Experiments}
As described in Section \ref{sec:dataset}, we use two dataset splits: \textit{generalization} and \textit{interpolation}. In the generalization split, the model is trained on all topologies except ITW and CHA, which are reserved for evaluation.  This experiment assesses how well the model can learn the influence of different zeolite frameworks on adsorption properties. The interpolation split, on the other hand, evaluates the model’s ability to capture the effect of different aluminium distributions on CO$_2$ adsorption. In both cases, we compare SymGNN against a standard GNN with identical hyperparameters, where the FiLM layer is replaced by a conventional linear layer.

All models are trained for 400 epochs, using the AdamW \cite{loshchilov2018decoupled} optimizer with default weights and a batch size of 64. The model is trained using mean-squared error loss for both the heat of adsorption and loading derivative. During training, edge dropout \cite{rong2020dropedge} with a probability of 0.5 is used to regularize the network. Due to the limited amount of isotherm data, the network is initially trained using only the heat of adsorption objective for the first 100 epochs. This approach mimics pre-training strategies used in fields like natural language processing (NLP), where models first learn general patterns before fine-tuning on specific tasks. This phase allows the model to establish the relationship between adsorption properties and framework geometry. In the following 25 epochs, the coefficient for the loading derivative loss is linearly increased from 0 to 1. For loading predictions, we evaluate at 100 logarithmically spaced pressures, ranging from 0.01 kPa to 10,000 kPa. A random window of 25 pressures for each structure is used to calculate the loss to reduce overfitting.

To construct a graph representation of a zeolite, we use a binary node encoding, where silicon is represented as 0 and aluminum as 1. Undirected edges are drawn between atoms within a radius of 8\AA, while ensuring periodic boundary conditions are respected. Each edge is further annotated with the Euclidean distance between the connected atoms. 

We calculate the generators for each atomic position within a given topology. Since the goal is to leverage symmetry operations to inform the GNN about the crystal geometry, atom types are not considered in the calculation. Including them would cause most structures to belong to the least symmetric space group, which would remove any geometric information the generators carry. To determine the generators, we first obtain space group information from the GENPOS program of BCS \cite{aroyo2006bilbao}, then algorithmically identify the generators for each atomic orbit within the topology.

\subsection{Structure Characterization}
In experimental settings, the precise atomic structure of a zeolite is often unknown. Determining key structural properties, such as the Si/Al ratio or the specific atomic arrangement within the unit cell, can provide valuable insights into a material's adsorption behavior. To address this, we employ an optimization-based approach to infer likely structures based on adsorption data.

We adopt a genetic algorithm (GA)-based approach, where the genes represent the Si and Al atom assignments within the framework \cite{petkovic2024graph}. The algorithm starts with an initial population of 200 candidate structures, initialized randomly. At each iteration, the top 25 structures (elite selection) are preserved, while mutations are applied to both the best 25 and the second-best 25 structures, resulting in 50 structures undergoing modifications per generation. Mutations include small-scale Si/Al swaps, full random permutations, and the addition or removal of a single Al atom. After filtering out duplicate and symmetrically equivalent structures (which are removed with a 90\% probability), the population is replenished to 200 candidates to maintain diversity. In total, the GA runs for 50 generations, following which we extract the 25 best performing structures.

The fitness function follows the approach from \citet{petkovic2024graph}, where candidates are evaluated based on their agreement with the experimental isotherm. In addition, the fitness function penalizes unnecesarily introducing aluminium atoms. However, to mitigate potential biases in the model, we introduce an additional term that explicitly evaluates how well the predicted isotherm captures the overall shape of the experimental data. This adjustment helps refine the search towards physically meaningful solutions.

To assess the model's performance in structure characterization, we apply this method to several experimental isotherms from the literature. Specifically, we consider two MFI \cite{dunne1996calorimetric}, two MOR \cite{delgado2006adsorption,kwon2022tailoring}, and one LTA4A\cite{martin2014insights} zeolite, with varying Si/Al ratios. We analyze how well the algorithm can recover the correct structural parameters from the adsorption data. For this experiment, we used the SymGNN model trained on the interpolation data split.

\section{Results}
\subsection{Model Performance}
\begin{figure*}[!]
\centering
  \includegraphics[width=.95\linewidth]{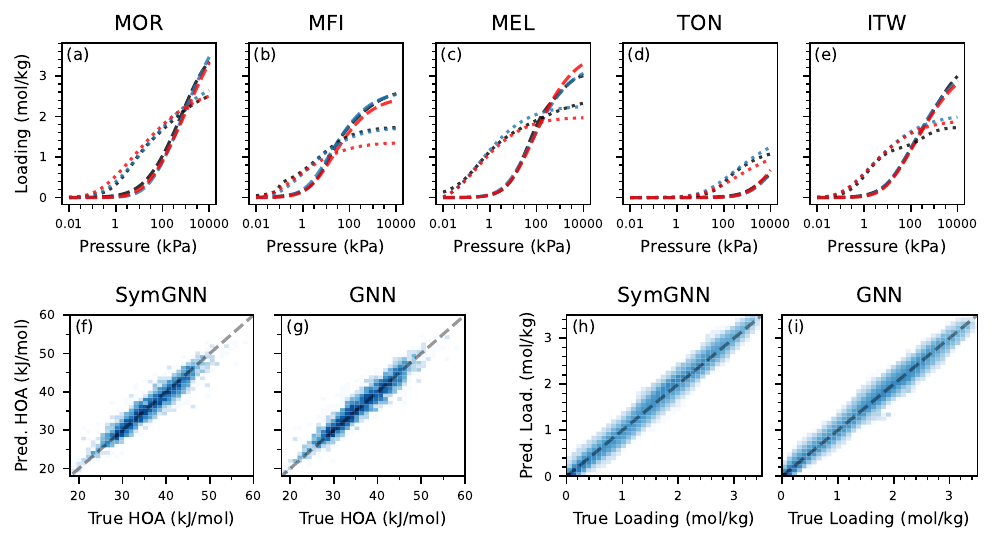}
  \caption{Comparison of SymGNN and regular GNN on the interpolation experiment. (a-e) True adsorption isotherms (black), SymGNN predicted isotherms (blue) and GNN predicted isotherms (red), for a high Si/Al ratio structure (dashed) and low Si/Al ratio structure (dotted) for each topology. (f,g) Parity plots for the heat of adsorption prediction. (h,i) Parity plots for the loading predictions. For all parity plots (f-i), darker blue indicates a higher count, and increases in log-scale.}
  \label{fig:intres}
\end{figure*}
To evaluate the performance of the different models in both interpolation and generalization experiments, we calculate the Mean Absolute Error (MAE) and Mean Squared Error (MSE) across various quantities. These include the heat of adsorption, the full adsorption isotherm, and the isotherm near saturation pressure (the final 10\% of the pressure range). The last metric provides insight into how well the model captures variations in loading caused by the framework structure and aluminium distribution. These metrics are summarized in Table \ref{tab:results}. 

\begin{table}[]
\small
  \caption{\ Performance of SymGNN and a regular GNN for both the generalization (gen) and interpolation (int) tasks.}
  \label{tab:results}
  \begin{tabular*}{0.48\textwidth}{@{\extracolsep{\fill}}lcccccc}
    \hline
    {} &  \multicolumn{2}{c}{Heat of adsorption} & \multicolumn{2}{c}{Isotherm} & \multicolumn{2}{c}{Isotherm sat.} \\
    \cmidrule(lr){2-3}
    \cmidrule(lr){4-5}
    \cmidrule(lr){6-7}
    {} & MAE & MSE & MAE & MSE & MAE & MSE \\
    
    \hline
    SymGNN (gen) & 1.64  & 4.75 & 0.35 & 0.19 & 0.16 & 0.04 \\
    GNN (gen) & 1.79 & 5.07 & 0.32 & 0.18 & 0.67 & 0.53 \\
    \hline
    SymGNN (int) & 1.42 & 3.72 & 0.07 & 0.01 & 0.10 & 0.02 \\
    GNN (int) & 1.45 & 3.87 & 0.07 & 0.01 & 0.10 & 0.02\\
    \hline
  \end{tabular*}
\end{table}

In the interpolation experiment, we observe no significant difference between the two models. Since both models have been trained on every topology present in the test set, the focus shifts away from the influence of the zeolite framework and more toward learning how aluminium distribution affects adsorption. As a result, explicitly modeling symmetries provides little additional benefit in this setting. This can be further seen in Figure \ref{fig:intres}, with both models having a similar performance in the prediction of the heat of adsorption and isotherm.

\begin{figure*}[]
\centering
  \includegraphics[width=.95\linewidth]{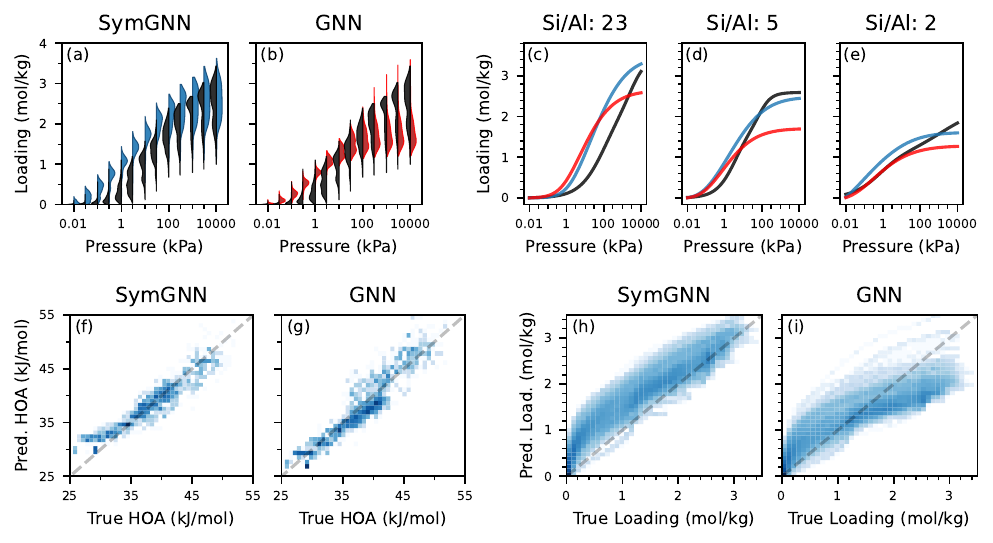}
  \caption{Comparison of SymGNN and regular GNN on the generalization experiment. (a,b) True loading (black) distribution at all simulated pressures compared with loading distribution obtained from SymGNN (blue) and GNN (red). (c-e) True adsorption isotherms (black), SymGNN predicted isotherms (blue) and GNN predicted isotherms (red) for ITW structures with varying Si/Al ratios. (f,g) Parity plots for the heat of adsorption prediction. (h,i) Parity plots for the loading predictions. For all parity plots (f-i), darker blue indicates a higher count, and increases in log-scale.}
  \label{fig:genres}
\end{figure*}

In contrast, the generalization experiment reveals a decline in performance for both models. However, the model that incorporates symmetry information achieves lower error in predicting the heat of adsorption and maintains comparable accuracy in predicting the isotherm near saturation pressure. To further analyze this, we compare the distributions of the true and predicted isotherms for both models, as shown in Figures \ref{fig:genres}a and \ref{fig:genres}b. The symmetry-informed model captures the overall behavior of the isotherm but increases the loading too early. In contrast, the standard graph neural network predicts isotherms with little variance, producing almost the same isotherm for each structure and severely underestimates the loading at higher pressures.

\begin{figure}[]
\centering
  \includegraphics[width=.95\linewidth]{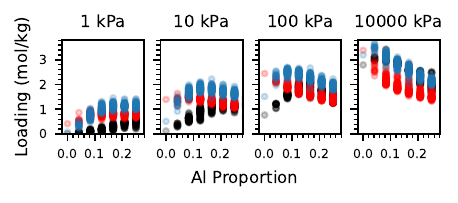}
  \caption{True (black) and predicted distribution of loading values as a function of the aluminium proportion for the symmetry informed (blue) and regular GNN (red) at different pressures.}
  \label{fig:isosctrpr}
\end{figure}

Parity plots for the heat of adsorption are shown in Figures \ref{fig:genres}f and \ref{fig:genres}g. For the SymGNN, we observe a slight overestimation of the heat of adsorption for lower values, whereas the regular GNN tends to understimate lower values, and overestimate higher values. Despite this, SymGNN successfully captures the underlying trends and generalizes well, effectively learning the influence of unseen zeolite topologies on the heat of adsorption.

In the parity plots for loading (Figures \ref{fig:genres}h and \ref{fig:genres}i), a distinct trend emerges. SymGNN primarily overestimates the loading, whereas the regular GNN overestimates lower loadings but underestimates higher ones. Examining the isotherm predictions for ITW structures with varying Si/Al ratios (Figures \ref{fig:genres}c–\ref{fig:genres}e), we find that SymGNN accurately captures the overall trend and the correct loading near saturation pressure. In contrast, the regular GNN increases the loading too early and fails to reach the correct saturation pressure. Additionally, SymGNN better captures the influence of aluminium distribution across different pressures (Figure \ref{fig:isosctrpr}), accurately modeling both the initial increase and subsequent decrease in loading, whereas the regular GNN only captures the decreasing trend. Overall, these results demonstrate that incorporating symmetry improves generalization to unseen zeolite structures, particularly in capturing adsorption trends across different frameworks, despite the model being trained on isotherms from only four other topologies.

\subsection{Symmetry Utilization Analysis}
While incorporating symmetry information into the model improves its performance, it is essential to determine whether the model has genuinely learned to leverage these symmetries or if the observed improvements arise from other factors. To this end, we examine whether the generator embedding network assigns distinct $\gamma$ and $\beta$ parameters to different sets of generators, indicating that the model differentiates between symmetry elements. Additionally, we analyze how the model’s predictions change when substituting the true generators of atoms in a zeolite with an alternative set, testing whether the learned symmetry representations meaningfully influence adsorption behavior. These experiments are carried out on the SymGNN model used in the generalization setting.

\begin{figure}[h!]
\centering
  \includegraphics[width=.95\linewidth]{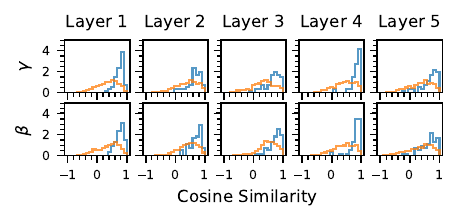}
  \caption{Cosine similiraty for $\gamma$ and $\beta$ parameters from similar generators (blue) and different generators (orange), for each message passing layer.}
  \label{fig:hypsym}
\end{figure}

In total, there are 61 unique sets of generators across all nodes in the dataset. To examine whether the model has learned distinct $\gamma$ and $\beta$ parameters for each unique set of generators, we calculate the cosine similarity between these parameters for different generators. Additionally, to assess whether the model has learned to associate similar generator sets with similar parameters, we define two distinct sets of generator pairs. The first set contains pairs $(i,j)$, where $G_i \subset G_j$ and $|G_j| - |G_i| = 1$, meaning one set of generators includes all elements of the other set, plus one additional generator. The second set contains pairs where this condition does not hold.

As shown in Figure \ref{fig:hypsym}, the model has indeed learned distinct $\gamma$ and $\beta$ parameters for the different sets of generators across all layers of the network. Moreover, similar generators tend to exhibit higher cosine similarity than dissimilar generators, suggesting that the model has learned a meaningful relationship between them. Additionally, the cosine similarity of the $\gamma$ and $\beta$ parameters for similar generators is statistically significantly greater than the cosine similarity for different generators, further confirming that the model has learned to associate similar generators with similar parameter values.

To analyze whether SymGNN bases its predictions on the geometric information provided by the generators, we replace the generators of the nodes in the test set (ITW and CHA), with the generators of nodes from a different zeolite. More specifically, for each orbit of nodes in both topologies, we replace their generators with the same generator from a different zeolite. For each generator replacement, we evaluate the model performance on the modified test set.

\begin{figure}[h]
\centering
  \includegraphics[width=.95\linewidth]{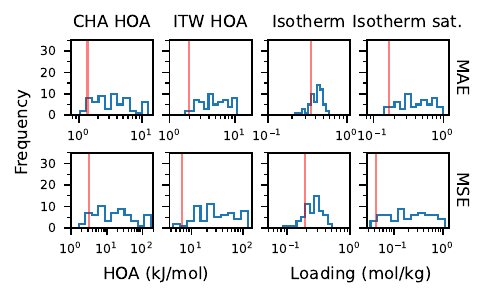}
  \caption{Distribution of evaluation metrics when replacing true generators of an orbit. The vertical red line indicates model performance when the original (correct) generators are used. Note that the x-axis is in log-scale.}
  \label{fig:wronggen}
\end{figure}

In Figure \ref{fig:wronggen}, we observe how the evaluation metrics are impacted when an incorrect set of generators is used for a given topology. Overall, the performance degrades significantly, rendering the model nearly unusable. While there are a few instances where the performance is marginally better, this is likely due to the use of a generator set that is similar to the correct one. In the full isotherm, there are more incorrect generators for which the error is lower, but this can be attributed to an inherent bias in our network when predicting isotherms, as the model performs notably better near saturation pressure. From this, we can conclude that the model indeed leverages the symmetries in the zeolite structures.

\subsection{Structure Characterization}

\begin{figure}[h!]
    \centering
    \includegraphics[width=0.95\linewidth]{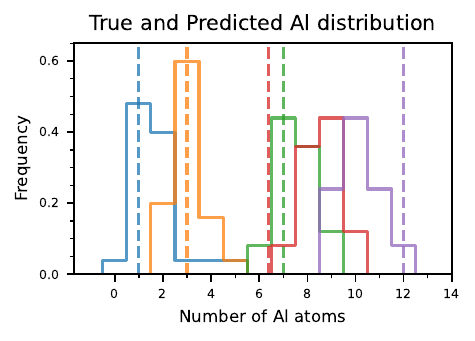}
    \caption{Aluminium distribution of experimental structures (dashed line) and aluminium distribution predicted by the genetic algorithm (histogram). Structures included are MFI with a Si/Al ratio of 95 (blue) and a Si/Al ratio of 31 (orange), MOR with a Si/Al ratio of 5.8 (green) and a Si/Al ratio of 6.5 (red) and LTA with a Si/Al ratio of 1 (purple). }
    \label{fig:ga_al}
\end{figure}

To assess our model's performance in structure characterization, we examine the aluminium distributions in the generated structures. Figure \ref{fig:ga_al} compares the predicted distribution of aluminium atoms per unit cell from our genetic algorithm with the true distribution. By generating a range of possible aluminium arrangements, our approach provides additional insight into the material, as real crystals often exhibit variations in their unit cell configurations. In the case of both MFI structures and one of the MOR structures, the predicted aluminium distribution is centered around the true value. However, for the other MOR structure, the model tends to overpredict the aluminium content, while for LTA4A, it underpredicts it. These deviations suggest that while the model captures key trends in aluminium placement, there is still room for improvement in accurately modeling specific cases.

As observed in the generalization experiment, the model struggles to fully generalize across different zeolite structures. While incorporating the isotherm shape into the fitness function improves performance, it may not completely resolve this limitation. A possible way forward is to increase the diversity of training data by incorporating more isotherms from a wider range of zeolite topologies. Additionally, fine-tuning the model using experimental data could enhance its ability to capture real-world adsorption behavior more accurately. Such improvements could make the model more reliable for structure characterization and broaden its applicability to new materials.

\section{Conclusion}
In this work, we introduced SymGNN, a symmetry-informed graph neural network capable of accurately predicting adsorption properties in zeolites. Our results demonstrate that incorporating structural information into message passing allows for improved generalization, enabling accurate predictions of both adsorption isotherms and heats of adsorption, even for unseen topologies. Despite being trained on a limited dataset, SymGNN exhibits strong predictive performance. The model effectively learns adsorption trends across different zeolite frameworks and Si/Al distributions, highlighting its robustness even when data is sparse. This makes it a promising approach for studying adsorption in materials where experimental data is limited.

A key finding of this work is that a model trained entirely on simulated isotherms can be used to analyze real zeolite structures. By applying SymGNN to experimental adsorption data, we demonstrated its potential for structure characterization, showing that it can infer properties such as the Si/Al ratio from adsorption trends. This suggests that machine learning models trained on computational data can bridge the gap to real-world applications.

One limitation of our study is the restricted availability of adsorption isotherms, both in terms of the number of samples and the diversity of zeolite topologies. While our model performs well across the available data, expanding the dataset to include more topologies and adsorption conditions would likely improve generalization further.

Looking ahead, generative models offer an exciting avenue for inverse design, allowing for the discovery of new zeolite structures with tailored adsorption properties. However, while such models have shown promise in MOFs \cite{fu2023mofdiff}, they only operate on a building block level. As such, their application at the atomic level for porous materials remains largely unexplored. Future work could explore how generative models can be combined with physics-informed learning to accelerate zeolite design.

Fine-tuning SymGNN with experimental data presents another promising direction. Incorporating real adsorption measurements into training could further improve both prediction accuracy and structure characterization, helping refine our understanding of real zeolite materials. This approach could also enhance the model’s ability to generalize beyond simulated conditions, making it even more applicable to practical adsorption studies.

Overall, this work highlights the potential of machine learning for adsorption modeling in nanoporous materials. By leveraging structured representations and data-driven learning, models like SymGNN provide a powerful tool for both predictive modeling and material characterization, paving the way for future advances in adsorption science and materials discovery.


\section*{Supporting Information}
Figure S1: Aluminium placement algorithms example; Figure S2: Number of samples per zeolite topology; Figure S3: Reduced simulation settings validation; Figure S4: RUPTURA validation; Table S1: Reduced simulation settings.

\section*{Acknowledgements}
This work used the Dutch national e-infrastructure with the support of the SURF Cooperative using grant no. EINF-10879. 

\section*{Author contributions}
Conceptualization M.P., V.M., J.M.V.L., S.C.; Data curation M.P., E.B.D.; Formal analysis M.P.; Funding acquisition V.M., S.C.; Investigation M.P.; Methodology M.P., J.M.V.L.; Project administration V.M., S.C; Software M.P., E.B.D.; Resources S.C.; Supervision J.M.V.L., V.M., S.C.; Validation M.P.; Visualization: M.P.; Writing – original draft: M.P.; Writing – review \& editing: M.P, E.B.D., J.M.V.L., V.M., S.C.
\section*{Competing interests}

The authors declare no competing interests.

\section*{Data availability}
Data for the zeolite structures and their adsorption properties, as well as the code for the models and experiments, is available at \url{https://doi.org/10.5281/zenodo.15085783}. Code for the PORRAN program is available at \url{https://doi.org/10.5281/zenodo.15050435}.

\bibliography{ms}





\renewcommand{\thesection}{S\Roman{section}}
\renewcommand{\thetable}{S\arabic{table}}  
\renewcommand{\thefigure}{S\arabic{figure}} 
\setcounter{figure}{0}
\setcounter{section}{0}
\setcounter{table}{0}

\onecolumngrid 

\newpage
\begin{center}
    \textbf{\Huge{Supporting Information}}

    \vspace{1.0cm}

    for

    \vspace{1.0cm}

    \textbf{\Large{Symmetry-Informed Graph Neural Networks for Carbon Dioxide Isotherm and Adsorption Prediction in Aluminum-Substituted Zeolites}}
\end{center}

\section{Zeolite structures}

\subsection{Aluminium placement algorithms}
To generate the dataset used in this project, the ZEORAN program \cite{romero2023adsorption} was used to place aluminium atoms in all-silica zeolites, using four algorithms for distributing the atoms throughout the structure. For some zeolite structures, the aluminium placement was done using PORRAN \cite{marko_petkovic_2025_15050436}, which is a Python extension of ZEORAN. The four algorithms include $clusters$, $chains$, $maximum$ $entropy$ and $random$. These algorithms make use of a graph representation of the zeolite, where edges between T-atoms are drawn if they are part of the same T-O-T bond.

The $clusters$ algorithm initially places an aluminium atom in a random position in the structure. Following this, the neighbours of aluminium atoms are recursively substituted with aluminium atoms, until the desired amount of substitutions is reached. Structures generated using this algorithm contain a high amount of non-L\"{o}wenstein bonds.

In the $chains$ algorithm, a user-defined number of chains (with a user-defined length per chain) of aluminium atoms is placed throughout the zeolite structures. Chains are placed in such a way that two separate chains do not connect. Furthermore, chains do not have branches, meaning that each aluminium atom in the chain will have one (if at the end) or two neighbours.

When using the $maximum$ $entropy$ algorithm, aluminium atoms are placed approximately uniformly distributed throughout the structure. In ZEORAN, this is achieved using a random walk. In PORRAN, aluminium atoms are iteratively placed, where silicon atoms which are the furthest from their closest aluminium atom have a proportionally higher chance to be selected. As such, there should be no non-L\"{o}wenstein bonds present in these structures.

Finally, the $random$ alogrithm randomly places aluminium atoms in the structure. Structures generated using this algorithm do not follow any particular distribution. 

\begin{figure}
    \centering
    \includegraphics[width=0.9\linewidth]{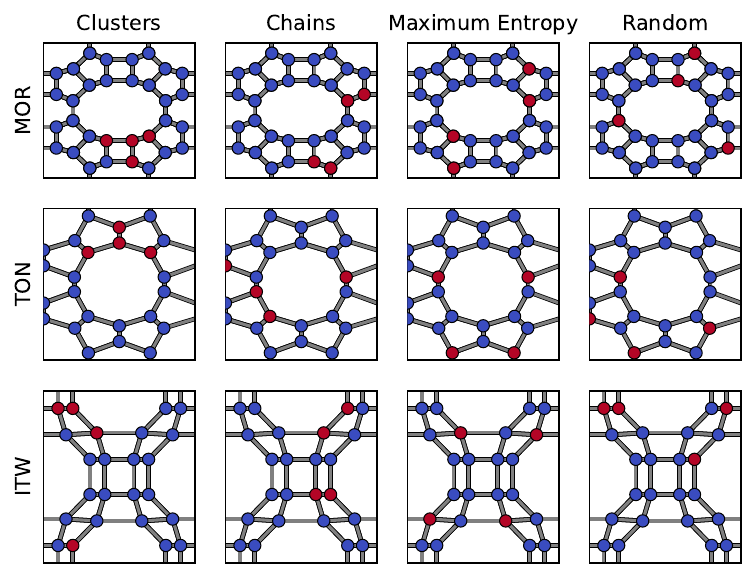}
    \caption{Example of structures generated with 4 aluminium substitutions for MOR, TON and ITW. In case of the $chains$ algorithm, 2 chains of length 2 were placed. Note that $clusters$ and $chains$ algorithms might place aluminium patterns crossing the periodic boundary. All structures are viewed along the z-axis.}
    \label{fig:enter-label}
\end{figure}

In Figure \ref{fig:enter-label}, each algorithm was used to generate an aluminium substitued strcuture for the MOR, TON and ITW zeolites. For each structure, 4 aluminium atoms were placed using their respective algorithms. 

\begin{figure}[]
    \centering
    \includegraphics[width=0.9\linewidth]{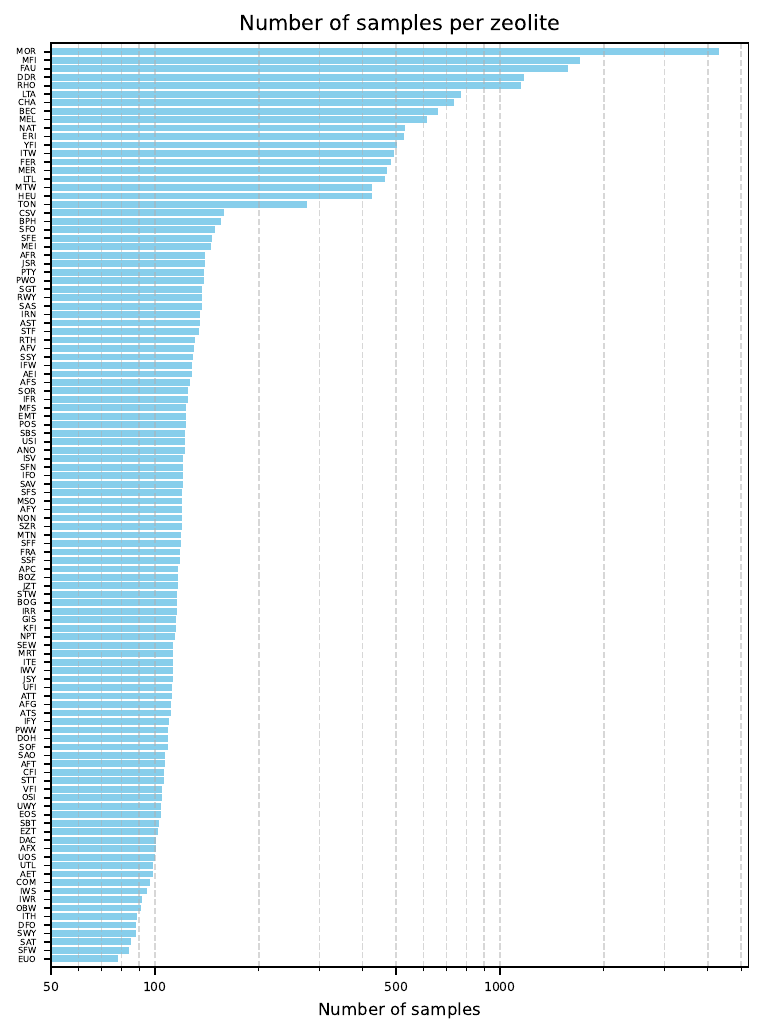}
    \caption{Number of samples for each zeolite topology. Note that the x-axis is in log-scale.}
    \label{fig:counts}
\end{figure}
\subsection{Zeolite topologies}
In Figure \ref{fig:counts}, the number of samples for each topology used in this work can be found. While more structures were originally simulated, structures with a heat of adsorption error of higher than 1.5 kJ/mol or an error higher than 5\% of the range of the heat of adsorption for that topology were dropped. Topologies for which too much data was dropped (less than 75 structures) as part of this filtering were not considered in this work.

\section{Isotherm simulation validation}
\subsection{Reduced simulation settings}

Due to the large number of simulations needed to obtain an isotherm for a single structure, generating a large and varied dataset can be relatively time consuming. To speed up these simulations, some approximations can be made, such as reducing the number of unit cells used. As a result of the unit cell reduction, the super cell used in the simulation might not have the size of two times the cuttoff range for the Lennard-Jones potential. Depending on the zeolite topology, this can lead to interactions between atoms not being modeled properly. Therefore, it is necesary to verify that a simulation with the reduced number of unit cells produces results that are in agreement with the simulation with the correct number of unit cells.

\begin{table}[h]
\centering
  \caption{Adsorption isotherm simulation settings for each topology}
  \label{tab:isosttngs}
  \begin{tabular*}{0.5\textwidth}{@{\extracolsep{\fill}}lcc}
    \hline
    {} & Full simulation box & Reduced simulation box \\
    {} & (\#unit cells) & (\#unit cells) \\
    \hline
    MOR & 2x2x4 & 1x1x2 \\
    MFI & 2x2x2 & 1x1x2 \\
    MEL & 2x2x2 & 1x1x2 \\
    TON & 2x2x5 & 2x2x5 \\
    ITW & 2x2x3 & 2x2x3 \\
    \hline
  \end{tabular*}
\end{table}

\begin{figure}[]
    \centering
    \includegraphics[width=0.95\linewidth]{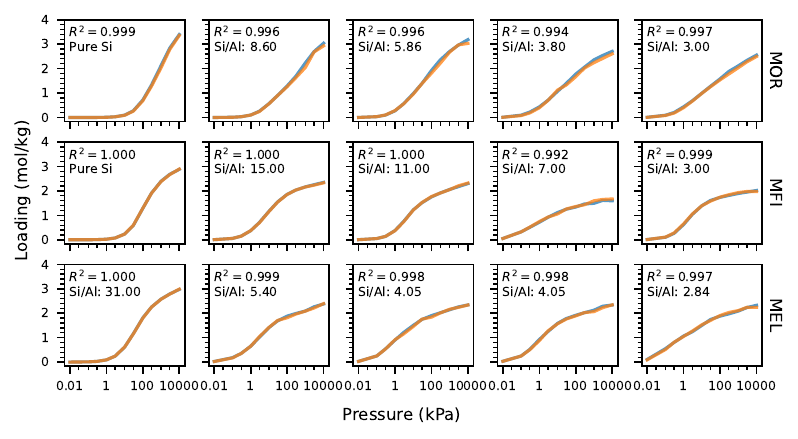}
    \caption{Isotherms for MOR, MFI and MEL structures with various Si/Al ratios, as predicted by the simulations using the full simulation box (blue), and the simulations using the reduced box (orange).}
    \label{fig:isocomp}
\end{figure}

The number of cycles and unit cells used for each zeolite topology can be found in Table \ref{tab:isosttngs}. For the MOR, MFI and MEL zeolite, the number of unit cells used was reduced. To verify that the full and reduced simulations are in agreement, we selected five configurations of each zeolite, with varying Si/Al ratios. For each of these structures, we carried out a simulation using both the full and reduced settings. A comparison between the two can be found in Figure \ref{fig:isocomp}. For all different structures, we notice that there is a near-perfect agreement between the two simulation settings, with only minor fluctuations in the reduced simulation. Therefore, we can shorten these simulations using the reduced settings without any significant sacrifices in accuracy.

\subsection{Isotherm Fitting}
\begin{figure}[]
    \centering
    \includegraphics[width=0.95\linewidth]{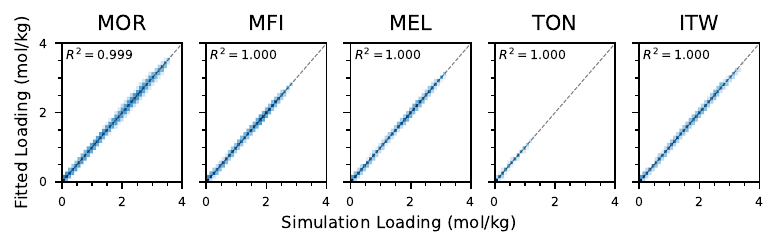}
    \caption{Parity plots for the isotherm fitting using RUPTURA. Darker blue indicates a higher count. Note that the color gets darker in log-scale.}
    \label{fig:ruptarity}
\end{figure}

Since we use reduced simulation settings for some of the isotherms simulations, certain fluctuations might occur in the simulated loading. In order to minimize the effect of these fluctuations, we fit the 2-site Langmuir-Freundlich equation on the simulated loadings, using RUPTURA \cite{sharma2023ruptura}. We compare the simulated loading with the fitted loading in Figure \ref{fig:ruptarity}. As can be seen, there are some minor deviations from the diagonal in the parity plots, which suggests some outliers have been smoothed out, while the overall correlation between the fitted and simulated isotherms is excellent. As such, we can conclude that using RUPTURA to fit the 2-site Langmuir-Freundlich does not affect the overall shape of the isotherm while smoothing out fluctuations.

\end{document}